**Unconventional Crystallization Pathway Bypassing the Intermediate Cubic Phase in Phase-change Superlattices**

*Bai-Qian Wang*[1], *Nian-Ke Chen*[1,*], *Yao-Jie Wang*[1], *Jia Sun*[1], *Yu-Ting Huang*[1], *Ming Xu*[2], *Shengbai Zhang*[3], *and Xian-Bin Li*[1,*]

[1]State Key Laboratory of Integrated Optoelectronics, College of Electronic Science and Engineering, Jilin University, 130012 Changchun, China
[2]School of Integrated Circuits, Huazhong University of Science and Technology, 430074 Wuhan, China
[3]Department of Physics, Applied Physics, and Astronomy, Rensselaer Polytechnic Institute, 12180 Troy, New York, USA
*Corresponding author, E-mail address: chennianke@jlu.edu.cn (N.-K. Chen) or lixianbin@jlu.edu.cn (X.-B. Li)

**Abstract**
The Ge-Sb-Te (GST) superlattice phase-change material is a promising candidate for overcoming the high power-consumption of phase-change memory (PCM). However, the working mechanism of the superlattice PCM remains controversial. Partial amorphization, which is currently considered the most plausible mechanism, remains hotly debated: how does the partially amorphized GST recrystallize into its superlattice phase instead of the conventionally expected cubic phase? Here, we address this issue using large-scale molecular dynamics simulations enabled by a machine-learning interatomic potential. Starting from a partially melted GST superlattice, we demonstrate that the residual crystalline regions serve as nuclei, enabling the amorphous GST to recrystallize directly into the superlattice phase without passing through the intermediate cubic phase. Moreover, the recrystallized phase is not an ideal superlattice, but rather a structurally ordered and chemically disordered defective superlattice characterized by antisite defects and stacking faults. The defective superlattice region is also more susceptible to melting than the defect-free superlattice, and thereby can act as the active region of the PCM device. These results help to clarify the longstanding debates concerning the mechanism of superlattice-based PCM.

## Introduction

The rapid development of big data and artificial intelligence technologies has posed severe challenges to conventional nonvolatile memory technologies (1-4). Phase-change memory (PCM) is a highly promising emerging nonvolatile memory technology that encodes information by utilizing the pronounced contrast in electrical or optical properties between the crystalline and amorphous states of phase-change materials (4-7). Although PCM is considered a key candidate for post-Moore-era memory technologies owing to its high-speed operation, long data retention, and high scalability, its high power-consumption remains a significant obstacle to practical applications (8-13). The development of superlattice phase-change materials provides an effective route to overcoming this challenge, owing to their significantly reduced power consumption compared with conventional phase-change materials (14-16). Despite extensive studies, the working mechanism of superlattice-based PCM remains a longstanding controversy in the community (17-19).

Although the mechanism of order-to-order phase transition has long been proposed and discussed (14, 20-24), the proposed models still suffer from limitations of low signal contrast, high energy barriers, or strong dependence on specific crystal configurations (19, 25-27). In recent years, growing experimental and computational evidence has suggested that partial amorphization occurs in superlattice phase-change materials (18, 28-32), with the formation of amorphous regions also being directly observed in recent electron microscopy experiments (28, 30).

However, despite these new findings, the mechanism of partial amorphization remains controversial, because it contradicts the previous widely accepted view that amorphous Ge-Sb-Te (GST) preferentially recrystallizes into a metastable cubic rock-salt (RS) phase with randomly distributed vacancies, rather than into a superlattice structure with ordered vacancies (33, 34). In other words, if the superlattice is amorphized during the RESET process, it would be expected to recrystallize into RS GST and thereafter behave like conventional RS-GST. This scenario is inconsistent with the experimentally demonstrated endurance of phase-change superlattices up to $10^8$ cycles (28). Therefore, to resolve this contradiction, it is essential to clarify whether the amorphous phase of superlattice phase-change materials can directly transform back into the superlattice or superlattice-like structure instead of recrystallizing into the conventional cubic phase.

Experimentally, in-situ characterization of phase-change superlattice materials near electrodes and tracking their crystallization kinetics are extremely difficult, leading to a lack of convincing experimental evidence. As for theoretical calculations, the simulation scale accessible to conventional density functional theory (DFT) calculations is too limited to directly model phase transitions involving sufficiently large crystalline-amorphous interfaces in the superlattices containing multiple van der Waals (vdW) gaps. Fortunately, machine-learning interatomic potentials (MLIP) can address such mesoscopic physical problems and have been successfully applied to large-scale simulations of phase-change materials in recent researches (35-39). MLIPs can accurately reproduce total energies and atomic forces, enable large-scale simulations while retain consistency with DFT-level accuracy (40-44).

In this paper, large-scale molecular dynamics (MD) simulations based on MLIP were performed to investigate the recrystallization of the partially amorphized superlattice phase-change materials. The MD simulations directly reproduced the recrystallization process, demonstrating that the amorphous phase can indeed transform back into the superlattice phase. During this process, the residual crystalline superlattice serves as a structural template for growth, significantly promoting recovery of the superlattice phase without the nucleation of the cubic phase. Furthermore, the simulations successfully reproduce the experimentally observed stacking faults and, for the first time, reveal their underlying atomic-scale formation mechanism. Finally, we found that the recrystallized superlattice phase is highly defective, containing numerous antisite defects. The defective superlattice phase can be regarded as a metastable phase resulting from rapid recrystallization. This defective phase makes it more susceptible to melting or amorphization, which in turn may help maintain the stability of the active region of partial amorphization during device cycling. These findings not only clarify the longstanding doubts regarding recrystallization into the superlattice phase, but also provide important clues toward understanding the working mechanism of superlattice phase-change materials.

## Results and discussion

### 1. Demonstration of the recrystallization of superlattice

To simulate the recrystallization of the partially amorphized superlattice phase-change materials, a partially amorphized model containing $Ge_2Sb_2Te_5$, $Ge_1Sb_2Te_4$ and $Sb_2Te_3$ sublayers, is firstly constructed by the melt-quench MD simulations, where part of the crystal is fixed to guarantee the coexistence of superlattice crystalline and amorphous phases. Figure 1a shows the structure of the constructed model. Then, the model is annealed at 700 K by MLIP-based MD to drive the recrystallization of the partially-amorphized superlattice. Figure 1a-f shows the snapshots of the structure during the recrystallization process. Obviously, the crystal-amorphous interface moved continuously, and the amorphous region gradually transformed into crystalline phase within 2.4 ns. To quantitatively characterize the recrystallization process, the amorphous region of the initial model (as indicated by the dashed lines in Figure 1g) is analyzed by the pair correlation function (PCF) and the deviation of bond-angle distribution from 90° ($D_{BAD}$) (29, 45, 46). The PCF is used to characterize long-range structural order, while the $D_{BAD}$ reflects local atomic disorder because the bond angles in the crystalline superlattice are predominantly close to 90°. The gradually emerging long-range peaks in the PCF shown in Figure 1h indicate a transition from the disordered amorphous phase to an ordered structure. After 2400 ps, the PCF of the recrystallized region agrees well with that of the crystalline superlattice. Meanwhile, the time evolution of the $D_{BAD}$ in the amorphous region (Figure 1i) shows a gradual decrease toward the value of the crystalline superlattice, indicating progressive recrystallization from the amorphous phase to the crystalline phase. According to the evolution of $D_{BAD}$, the recrystallization is nearly completed within 600 ps.

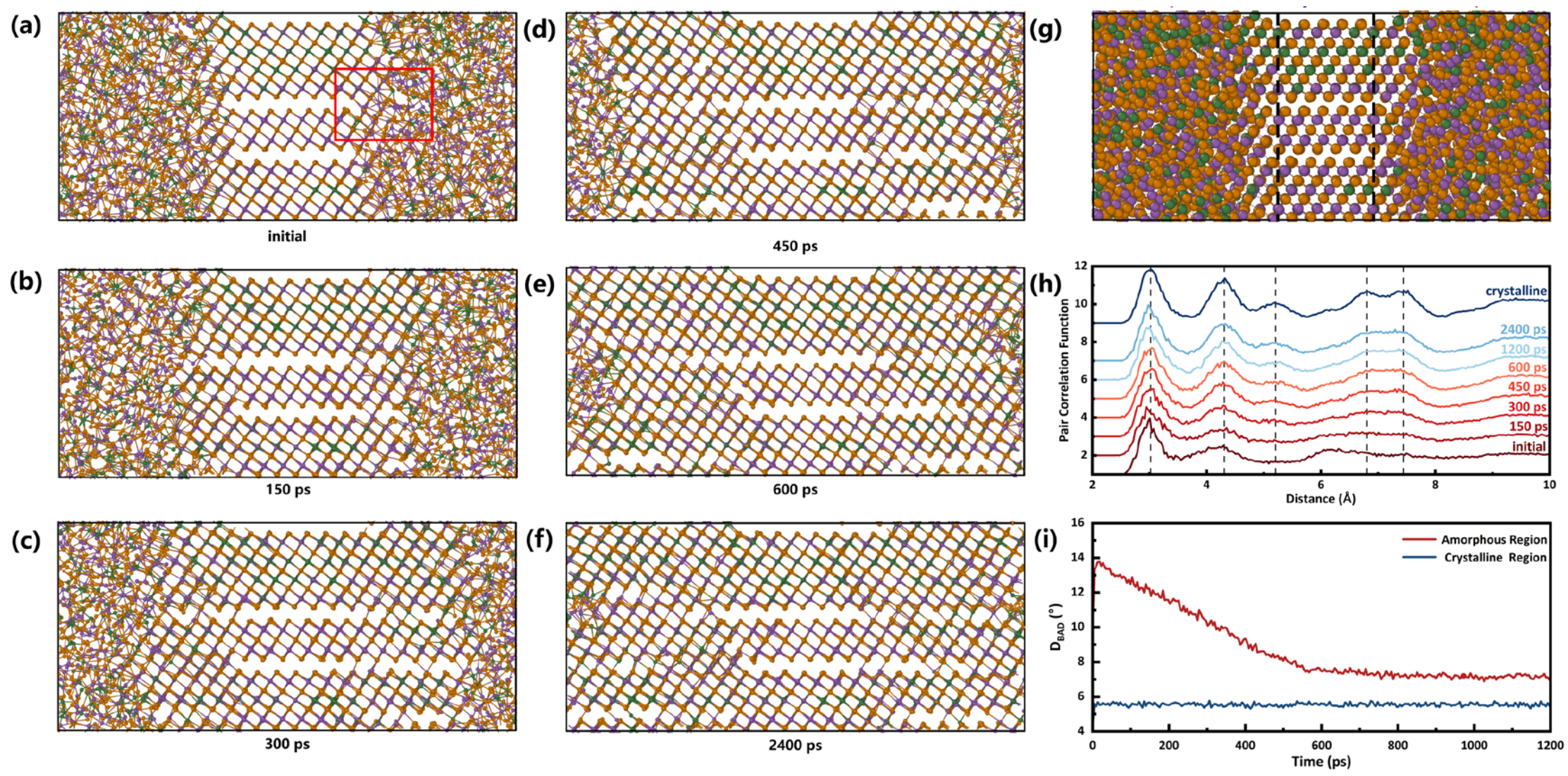

**Fig. 1 | The recrystallization of the partially amorphized Ge-Sb-Te/$Sb_2Te_3$ superlattice. a** The partially amorphized structure obtained by melt-quenching method in MLIP-MD, which is the initial structure of the recrystallization process. The green, purple, and orange balls represent Ge, Sb, and Te atoms, respectively. **b-f** Snapshots of the structures during the crystallization process at 700 K. **g** The crystalline and amorphous regions of the initial structure which are divided by the dash line. **h** The pair correlation function (PCF) of the crystalline and amorphous regions at different time during the crystallization process. **i** The deviation of bond angle distribution from 90° ($D_{BAD}$) of the crystalline and amorphous regions during the crystallization process. The $D_{BAD}$ of the crystalline superlattice at 700 K is about 5.5° which is the result of thermal fluctuations.

Note that, the $D_{BAD}$ of the amorphous region after crystallization is slightly higher than that of the ideal crystalline superlattice, indicating structural differences between the recrystallized phase and the perfect superlattice. This discrepancy arises from the presence of numerous defects in the recrystallized phase, including antisite $Ge_{Te}$, $Sb_{Te}$, $Te_{Ge}$ or $Te_{Sb}$ defects and stacking faults. These defects can directly be observed in the snapshots of recrystallized structures (see Figure 1f). Notably, the stacking faults reproduced in the present simulations are consistent with experimental observations, whose formation mechanism remained poorly understood (47, 48). Despite the presence of these defects, we emphasize that the vdW gaps are indeed reconstructed in the recrystallized phase. The vdW gaps are not clearly visible in Figure 1f because the presence of antisite defects causes some Ge and Sb atoms to occupy the vdW-gap regions. In brief, the recrystallization of partially amorphized superlattice leads to the formation of an ordered structure with layered characteristics, containing defects and stacking faults.

To further confirm that the recrystallized phase is a superlattice structure rather than the conventional cubic phase, X-ray diffraction (XRD) simulations of perfect GST superlattice (GST-SL), rock-sault GST (RS-GST), and the recrystallized phase were performed and compared. To properly include the influence of thermal fluctuations, XRD simulations were performed using the GST-SL and RS-GST structures (Figures 2a and b) annealed at 700 K. As shown in Figure 2c, the XRD peaks of the recrystallized model exhibit a high degree of similarity to those of perfect GST-SL rather than RS-GST. The time evolution of the XRD patterns from amorphous to crystalline phases, further demonstrating that the XRD results reliably capture the structure of the recrystallized phase, rather than being solely determined by crystalline seeds. Therefore, the recrystallized phase is essentially a superlattice.

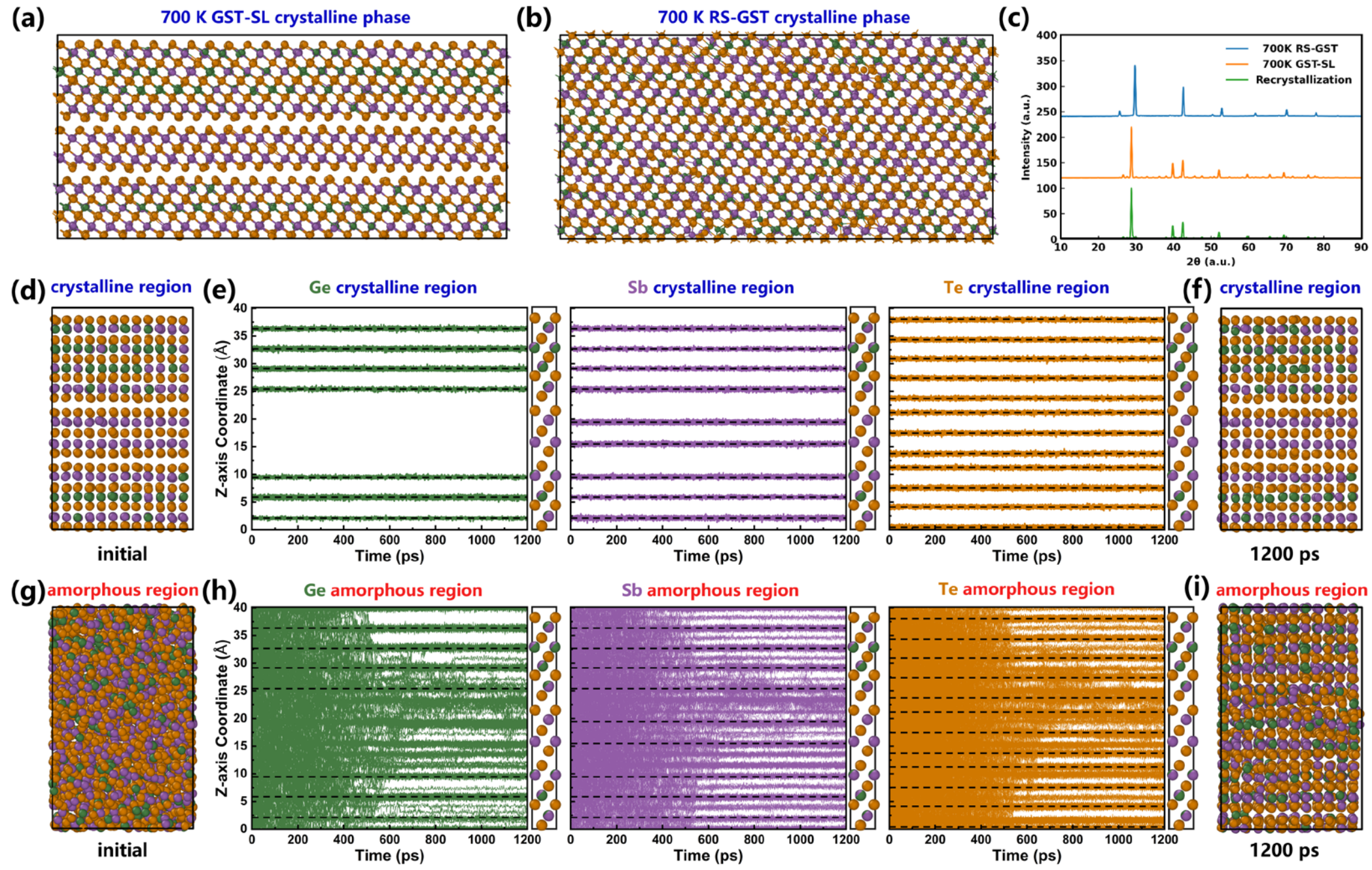

**Fig. 2 | Structural characteristics of the recrystallized structure. a** and **b** present the perfect GST-SL and RS-GST structures at 700 K. **c.** The comparison of XRD patterns of the recrystallized model, perfect GST-SL and RS-GST. **d.** Initial side view of the crystalline region. **e.** The atomic trajectories of the Ge, Sb and Te atoms in the crystalline region during the MD simulations. A schematic superlattice unit cell is presented for comparison, where the mixed colors on atoms represent the intermixing of Ge and Sb atoms. **f.** Side view of the crystalline region at 1200 ps of the MD simulations. **g.** Initial side view of the amorphous region. **h.** The atomic trajectories of the Ge, Sb and Te atoms in the amorphous region during the MD simulations. The dashed lines indicate the positions of atoms in the perfect crystalline phase. **f.** Side view of the amorphous region at 1200 ps of the MD simulations.

To capture the atomic picture of the recrystallization process, the MD trajectories of Ge, Sb and Te atoms are extracted and projected onto z axis (Figure 2d-i). Figures 2d-f and Figures 2g-i show atomic trajectories in crystalline and amorphous regions (as indicated in Figure 1g) with cross-section view, respectively. The corresponding atom positions of the perfect superlattice crystal structure are shown on the right for comparison. For the crystalline region, distinct sublayers of $Ge_2Sb_2Te_5$, $Ge_1Sb_2Te_4$ with Ge/Sb intermixing (49, 50) and $Sb_2Te_3$ in the superlattice can be seen throughout the MD simulations. The interlayer spacing along the z-direction between Te atomic layers across the vdW gap is smaller than that between Te layers within the lattice, indicating a clear vdW gap.

For the amorphous region, the time evolution of the z-axis coordinates of Ge, Sb, and Te atoms during the MD simulations clearly captures the transition from a disordered to an ordered structure (Figure 2h). The dashed lines in Figure 2h represent the ideal atomic positions in the perfect superlattice crystalline phase. After crystallization, most atoms in the amorphous region are incorporated into the correct lattice sites of the crystalline superlattice (aligned with the dashed lines), whereas a fraction of atoms occupy incorrect sites (deviating from the dashed lines), leading to defect formation. Atomic misalignment along the z direction, where Ge/Sb atoms are found in Te layers or Te atoms in Ge/Sb layers, indicates the formation of antisite defects and stacking faults (see a further discussion later). In addition, due to the limited recrystallization time, the $Sb_2Te_3$ layer is not recovered, leading to the formation of new Ge-Sb-Te sublayers.

**2. Atomic mechanism of the recrystallization**

During the crystallization process, the migration of the crystal-amorphous interface and the formation of vdW interfaces are crucial for the formation of the superlattice structure. Figure 3 illustrates atomic rearrangements at the vdW interface during crystallization, with highlighted atoms helping to visualize the process. Figure 3a shows the initial structure of the interface region indicated by the red frame in Figure 1a. Figures 3b-e illustrate the continuous movement of the crystalline-amorphous interface, indicating a growth-driven process. Here, the residual crystalline-phase region serves as the template of the growth. However, the disordered atoms do not transform into a highly ordered lattice in a single step. The newly formed lattice near the interface is still distorted. Homobonds, such as Ge-Ge, Sb-Sb, Ge-Sb and Te-Te bonds (also called wrong bonds), also exist in the interface. By carefully examining the evolution of the interface, we deduce the following growth rule: (1) Atoms in the amorphous phase generally exhibit relatively low coordination numbers and tend to evolve toward a crystalline lattice with higher coordination. (2) While the formation of normal Ge-Te and Sb-Te bonds is energetically preferred, “wrong” bonds in a higher-coordinated environment may also contribute to energy reduction comparing to the amorphous state. (3) Growth is primarily governed by nearest-neighbor bonding; thus, once an atom occupies an incorrect site (e.g., forming an antisite defect), it subsequently affects the local bonding environment and atomic arrangement. Based on this rule, the atomic-scale mechanisms underlying the formation of antisite defects and stacking faults can also be understood. To clearly illustrate this process, a schematic diagram is used to describe the growth mechanism (as shown in Figure 4).

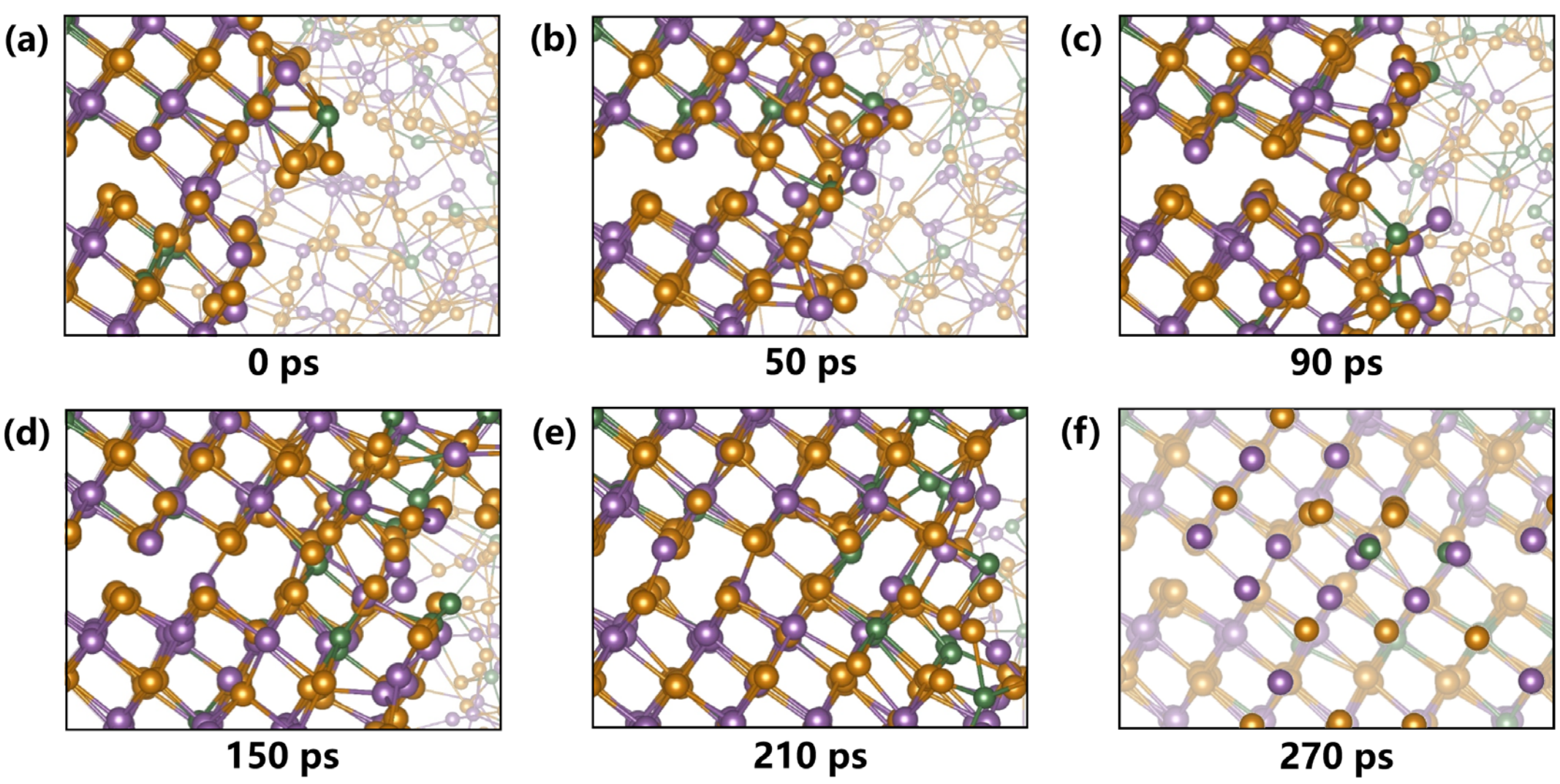

**Fig. 3 | Crystal growth process at the crystalline-amorphous interface. a-f** Snapshots illustrating the evolution of the crystalline-amorphous interface during the recrystallization process in a selected region (indicated by the red frame in Figure 1a). Recrystallized atoms and boundary atoms are highlighted, while atoms inside the amorphous region are smaller and shaded in gray. The antisite defects formed after the growth process are highlighted in **f**.

Figure 4a shows four typical types of local atomic configurations in the superlattice phase-change materials. As cations, both Ge and Sb atoms adopt octahedral (sixfold) coordination with Te atoms as nearest neighbors. In contrast, the local environment of Te atoms can be divided into two cases. Inside the layer block, Te atoms maintain sixfold coordination with neighboring Ge or Sb atoms. Near the vdW interface, each Te atom bonds with three cations and simultaneously interacts via van der Waals forces with other three Te atoms across the vdW gap. Figure 4b illustrates the formation of the perfect superlattice during the growth process. With the existing crystalline region acting as the template, atoms inside layer blocks arrange according to short-range coordination through normal bonding, while Te atoms across the vdW boundary are stabilized via vdW interactions. In this way, the medium- and long-range order is gradually established, resulting in the formation of an ideally ordered superlattice. Figure 4c illustrates the generation of antisite defects nearby the vdW gaps, which is frequently observed in our MD simulations. As discussed above, formation of a highly coordinated lattice is energetically favored. Therefore, "wrong" bonds may also form during the growth process, giving rise to the formation of antisite defects. Figure 4d illustrates the formation of a stacking fault. In this case, the upper and lower layer blocks exhibit different lateral growth rates—for example, when the upper block grows faster than the lower one—the in-plane growth on one side may complete earlier. Then, the out-of-plane growth may happen because the normal bonding of Ge-Te and Sb-Te is energetically favorable compared with the Te-Te vdW interactions. Consequently, the vdW gap will shift from its original layer to adjacent atomic layer, resulting in the formation of a stacking fault.

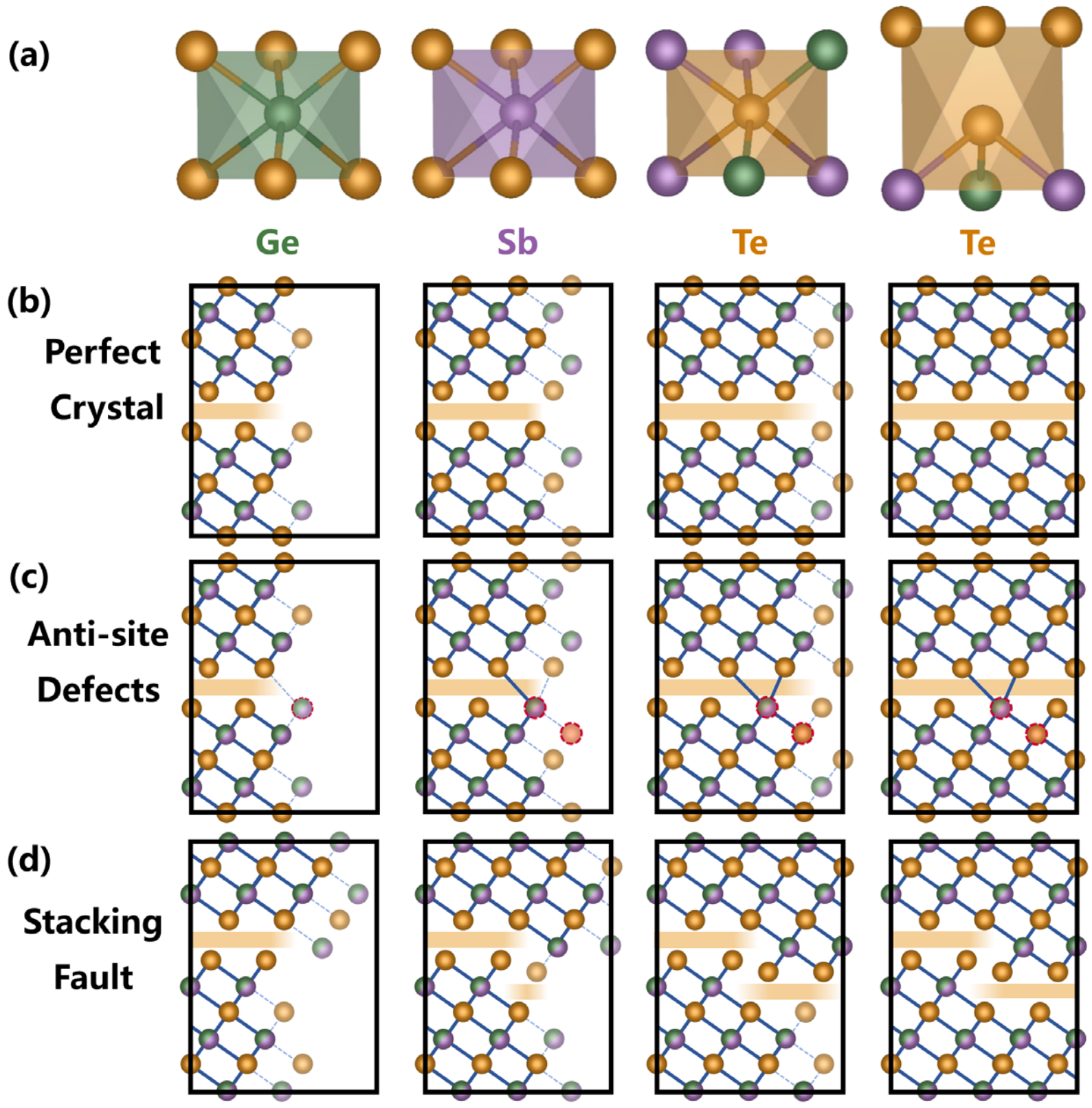

**Fig. 4 | Schematic illustration of the atomic-scale mechanism of the growth process. a.** Four typical types of local atomic configurations in perfect superlattice phase-change materials. **b.** The growth process of a perfect superlattice. **c.** The growth process of superlattice with the formation of antisite defects. **d.** The growth process of superlattice with the formation of stacking faults. The vdW gaps are highlighted by orange bars.

**3. The identification and impact of the defective superlattice.**

To further analyze the structural characteristics of the recrystallized phase, we use the smooth overlap of atomic positions (SOAP) based kernel similarity function to evaluate the similarity between the recrystallized phase and the perfect superlattice phase (see method for more details). Figures 5a-f show the kernel function values of all atoms during the growth process. The SOAP value of each atom is indicated by different colors, while the distributions of the kernel function value (probability density) are represented by the black curves in the right panels. We found that the distribution of the SOAP can be approximately decomposed into three peaks. Then, the distribution curves are fitted using three skew-normal distributions, as represented by the three dashed lines. Interestingly, the three skew-normal distributions can be attributed to the amorphous phase (green lines), perfect superlattice (blue lines), and the recrystallized phase (red lines). The three different phases can also be directly viewed via the color coding of atoms. As can be seen, green atoms are mostly concentrated in the uncrystallized amorphous region, blue atoms are localized in the nearly perfect superlattice, and red atoms represent the recrystallized superlattice with defects (namely, defective superlattice). As such, the atoms can be roughly divided into three categories: amorphous phase, perfect superlattice, and defective superlattice.

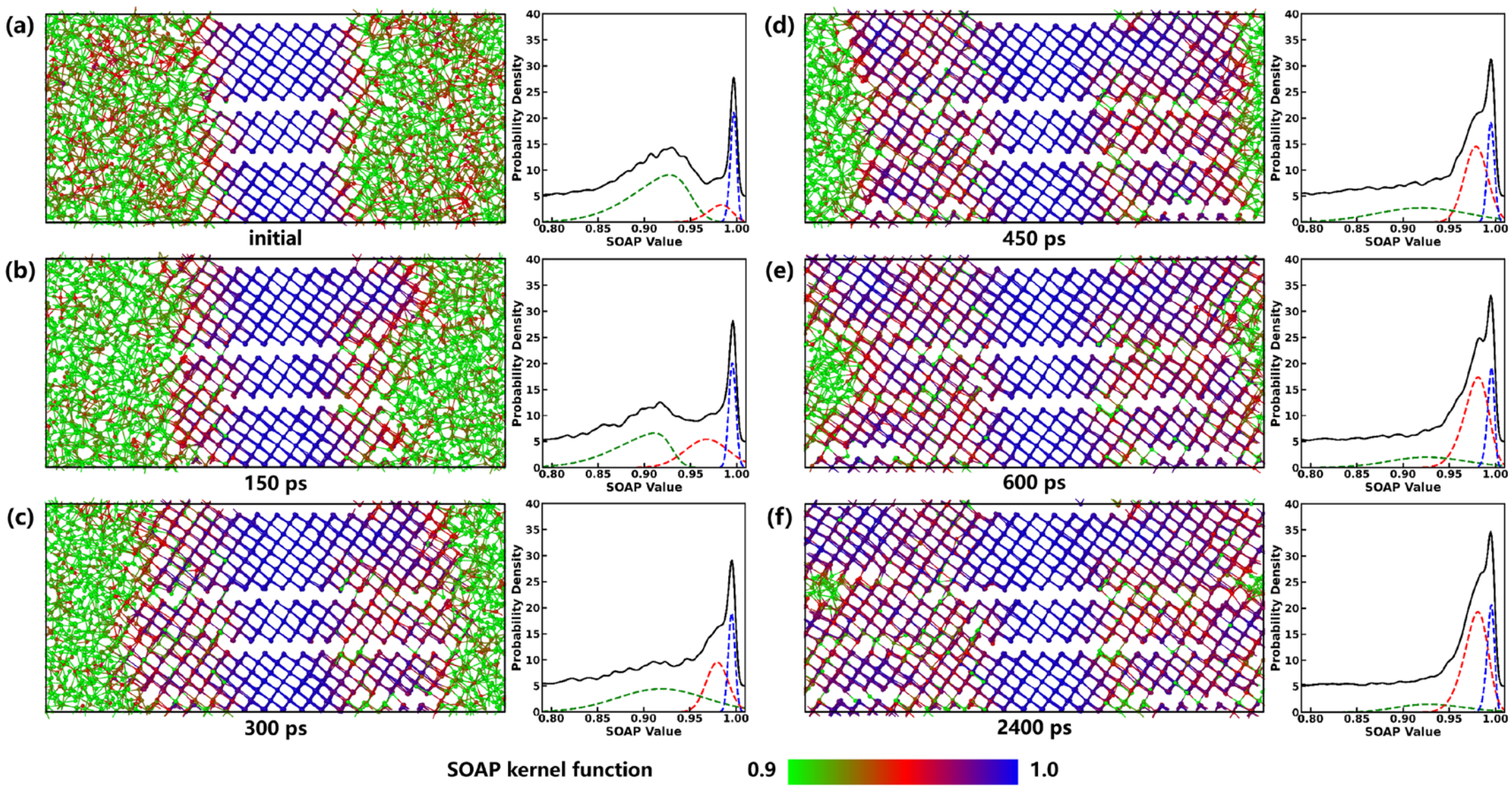

**Fig. 5 | Identification of the defective superlattice based on SOAP kernel similarity function. a-f** The SOAP characterization of the recrystallized model at different times of the MD simulations. Left panels: Atom-resolved SOAP values in the model, represented by color coding. Right panels: Distribution of the SOAP kernel function (black line) fitted with three skew-normal distributions, which can be attributed to the amorphous phase (green lines), perfect superlattice (blue lines), and the recrystallized phase (red lines). The black curve is shifted upward by 5 for clarity. The kernel function is computed with a reference to the local environments of atoms in the perfect superlattice crystal at 700 K.

The defective superlattice phase formed by this recrystallization will play an important role in the working process of PCM devices. Previous studies have reported that the melting of perfect superlattices is initiated by the formation of defects (29). Here, the presence of defects and stacking faults in the defective superlattice phase will make the structure more susceptible to melting than defect-free superlattices at the same temperature. To further confirm this deduction, the recrystallized model shown in Figure 6a was reheated to 1100 K for 9 ps. As shown in Figures 6b-d, the defective superlattice region first undergoes remelting, while the defect-free superlattice region remains relatively stable under the same conditions. Therefore, the defective superlattice, as an intermediate metastable phase formed during rapid recrystallization, is also more prone to melting than the perfect superlattice phase-change materials and thereby serve as the active region for partial amorphization in PCM superlattice devices to achieve low power consumption and high endurance.

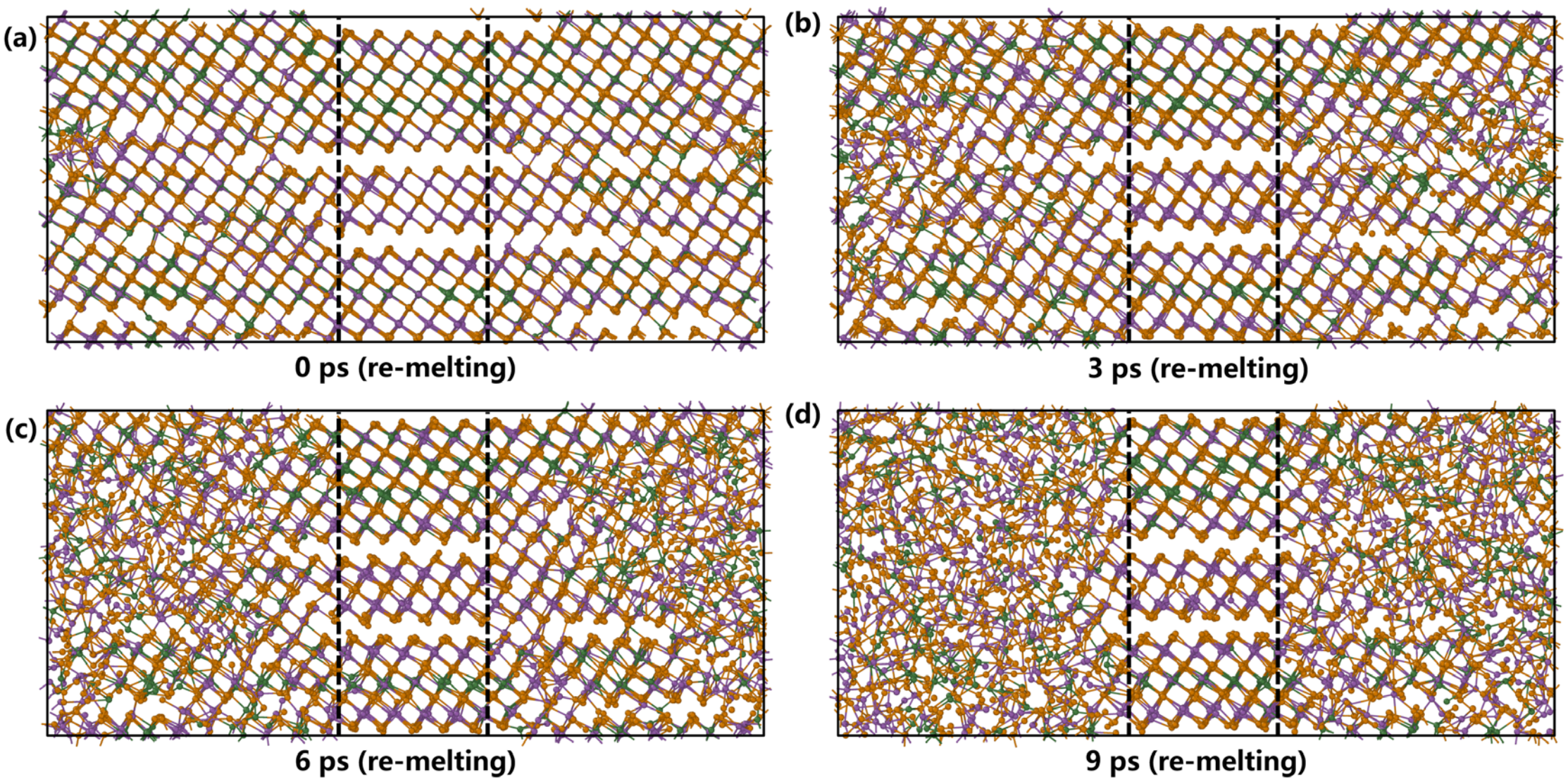

**Fig. 6 | The re-melting of the defective superlattice phase at 1100 K. a-d.** Snapshots of the model at different times during the remelting process. The dashed lines distinguish the central defect-free crystalline region from the recrystallized defective superlattice region.

## Conclusion

MLIP-MD simulations reveal that the partially amorphized region of the GST superlattice can directly transform into a defective superlattice phase, without undergoing the conventionally assumed cubic phase. This is because the surrounding crystalline region can act as a critical template for superlattice crystal growth. Due to the fast speed of growth process, the recrystallized phase is, in fact, a metastable defective superlattice with antisite defects and stacking faults. By carefully examine the evolution of the amorphous-crystalline interface, we conclude several rules of the growth process, according to which we further identified the formation mechanisms of the antisite defects, and stacking faults. In addition, this defective superlattice region is more susceptible to melting than the ideal superlattice phase-change material. Therefore, the defective superlattice phase may serve as the active region in superlattice-based PCM devices. These findings not only clarify the longstanding doubts regarding the recrystallization of the superlattice phase, but also provide important clues toward understanding the critical working mechanism of superlattice phase-change materials for future advanced memory and AI hardware.

## Method

The large-scale molecular dynamics simulations in this study were performed using LAMMPS (51), with a time step of 3 fs. Before the molecular dynamic simulations, structural relaxation was performed until the force convergence reached below 0.02 eV/Å. The molecular dynamics simulations were performed using the canonical ensemble (NVT ensemble), with the temperature controlled using the Nose-Hoover thermostat. Prior to the heating simulations, the system was first equilibrated at 300 K to minimize the influence of possible initial nonequilibrium effects on the subsequent heating process and dynamical analysis. The machine learning potential used in this study is the Ge-Sb-Te interatomic potential based on the Gaussian Approximation Potential (GAP) framework (52), published in 2023 by Zhou et al (53). Its training set covers the elemental substances and common binary phases in the Ge-Sb-Te system, and also fully considers the different phases and density variations of Ge-Sb-Te alloy, which gives it broad applicability for describing the systems investigated in this study.

In this study, the smooth overlap of atomic positions (SOAP) (54) was used to construct atomic descriptors for crystalline structures and to quantify the similarity between local atomic environments and a crystalline reference (55). For example, to quantify the similarity between the local environments of atoms indexed by $i$ and $j$, the SOAP kernel can be calculated as

$$k_{ij} = \widehat{\mathbf{P}}_i \cdot \widehat{\mathbf{P}}_j$$

where $\widehat{\mathbf{P}}_i$ and $\widehat{\mathbf{P}}_j$ are the normalized SOAP power spectrum vectors of the local environments of the two atoms, respectively. Here, $k_{ij}$ is the value of the SOAP kernel, which reflects the similarity between the two local environments. A value closer to 1 indicates that the local environment of the target atom is more similar to the crystalline state, whereas a smaller value suggests a more disordered local environment. The similarity of each atomic environment in a structure was evaluated using the SOAP kernel by comparing a target atom with atoms of the same species in the crystalline reference taken from molecular dynamics simulations of the superlattice crystal at 700 K. Specifically, the normalized SOAP power spectrum of the target atom was dotted with that of atoms of the same species in the crystal, and the maximum value of the resulting inner products was taken as the similarity between the local atomic environment of the target atom and that of atoms in the crystal.

In this work, structural ordering is evaluated using the deviation of bond-angle distribution ($D_{BAD}$), which quantifies the extent of structural distortion. In phase-change materials, the ideal angle between chemical bonds is close to 90°, whereas thermal effects and structural disorder can cause deviations from this ideal value. The deviation of bond-angle distribution for the bond angles in the structure is calculated as follows.

$$D_{BAD} = \sqrt{\frac{\sum_i (\theta_i - 90°)^2}{N_i}}$$

Here, $\theta_i \epsilon [0°, 135°]$ denotes the bond angles close to 90° in the structure, and $N_i$ is the number of bond angles that satisfy this criterion.

To analyze and compare the structural characteristics of the recrystallized phases, simulated powder X-ray diffraction patterns were generated using the *XRDCalculator* module implemented in the Python Materials Genomics (pymatgen) package (56). To facilitate direct comparison of the simulated diffraction patterns, the discrete XRD peaks were converted into continuous spectra by Gaussian broadening.

**Notes**

The authors declare no competing financial interests.

**Acknowledgments**

This work was supported by the National Science and Technology Major Project (Grant No. 2022ZD0117600), the National Natural Science Foundation of China (Grants No. 12274172, 12274180, 12504075), the Science and Technology Development Plan Project of Changchun, China (Grant No. 2024GZZ07), and the Fundamental Research Funds for the Central Universities. The High-Performance Computing Center (HPCC) at Jilin University for computational resources is also acknowledged.